\documentclass[11pt]{article}
\pdfoutput=1
\usepackage{jheppub}
\usepackage[T1]{fontenc} 

\usepackage{amssymb}
\usepackage{amsfonts}
\usepackage{amsbsy}
\usepackage{amsmath}
\usepackage{amsthm}
\usepackage{graphicx}
\usepackage{epstopdf}
\usepackage{multirow}
\usepackage{latexsym}
\usepackage{array}

\input cyracc.def 
\newfont{\twelvecyr}{wncyr10 at 12pt}
\def\sha{\text{\twelvecyr\cyracc{Sh}}}

\def\Z{\mathbb{Z}}
\def\F{\mathbb{F}}

\def\C{\mathbb{C}}

\def\P{\mathbb{P}}

\def\n3a{t}

\def\ge{{\mathfrak{e}}}

\def\gsu{{\mathfrak{su}}}

\def\gg{{\mathfrak{g}}}

\title{Sections, multisections, and $U(1)$ fields in F-theory}

\author[1]{David R.  Morrison}
\author[]{and}
\author[2]{Washington Taylor}

\affiliation[1]{Departments of Mathematics and  
Physics\\ University of California, Santa Barbara\\ Santa Barbara, CA 93106, USA}
\affiliation[2]{Center for Theoretical Physics\\
Department of Physics\\
Massachusetts Institute of Technology\\
77 Massachusetts Avenue\\
Cambridge, MA 02139, USA}

\emailAdd{{\tt drm} {\rm at} {\tt math.ucsb.edu}}
\emailAdd{{\tt wati} {\rm at} {\tt mit.edu}}

\preprint{UCSB Math  2014-12, MIT-CTP-4540}

\abstract{
We show that genus-one fibrations lacking a global section fit
naturally into the geometric moduli space of Weierstrass models.
Elliptic fibrations with multiple sections (nonzero Mordell-Weil
rank), which give rise in F-theory to abelian $U(1)$ fields, arise as
a subspace of the set of genus-one fibrations with multisections.
Higgsing of certain matter multiplets charged under abelian gauge
fields in the corresponding supergravity theories break the $U(1)$
gauge symmetry to a discrete gauge symmetry group.  We demonstrate
these results explicitly in the case of bisections, and describe the
general framework for multisections of higher degree.  We further show
that nearly every $U(1)$ gauge symmetry arising in an F-theory model can be
``unHiggsed'' to an $SU(2)$ gauge symmetry with adjoint matter, though
in certain situations this leads to a model in which a
superconformal field theory is coupled to a conventional gauge and gravity
theory.  The only exceptions are cases in which  the attempted
unHiggsing leads to a boundary point at an infinite distance from
the interior of the moduli space.}

\begin{document}

\maketitle

\flushbottom

\section{Introduction}

F-theory \cite{Vafa-F-theory,Morrison-Vafa-I,Morrison-Vafa-II} is a
nonperturbative approach to string theory in which the axiodilaton
$\tau=\chi + i e^{- \phi}$ of type IIB supergravity is specified by means of
an auxiliary complex torus (elliptic curve), and 7-branes serve as
sources for the RR scalar, providing an opportunity for 
$SL(2,\mathbb{Z})$-multivaluedness
of the $\tau$ field. 
In most work to date,
F-theory is compactified on a base $B_n$ of complex dimension $n$,
where the tori $\mathbb{C}/\langle1,\tau (\xi_1, \ldots, \xi_n)\rangle$ 
parameterized by
coordinates $\xi_i$ on the base are assumed to fit together to form a
Calabi-Yau $(n+1)$-fold $X_{n+1}$ that is elliptically fibered with
section, $\pi: X_{n+1}\rightarrow B_n$, so that 
(after appropriate blowing down) $X_{n+1}$ can be
described by a Weierstrass model
\begin{equation}
 y^2 = x^3+ f x + g \,,
\label{eq:Weierstrass}
\end{equation}
where $f, g$ are sections of line bundles ${\cal  O}(-4K), {\cal
  O}(-6K)$ on the base $B_n$ (locally described simply as functions of
the base coordinates).  The $7$-branes are located at the {\em
discriminant locus}\/ $\{4f^3+27g^2=0\}$, in a manner specified
by the Kodaira--N\'eron classification of singular fibers 
\cite{Kodaira,Neron}.

Recently, Braun and Morrison \cite{Braun-Morrison} considered a more
general class of F-theory compactification spaces, where the space
$X_{n+1}$ has a genus-one (torus) fibration, but no global section.
They identified a large number of examples of such genus-one
fibrations over the base $B_2 = \P^2$ in the comprehensive list
compiled by Kreuzer and Skarke \cite{Kreuzer-Skarke} of Calabi-Yau
threefolds that are hypersurfaces in toric varieties.  Any such
$X_{n+1}$ has a {\em Jacobian fibration}\/ $J_{n+1}$, which is an
elliptically fibered Calabi--Yau with section\footnote{This statement
  has been mathematically proven only for $n+1\le3$
  \cite{Kodaira,Gross,Nakayama-global}, but is likely true in
  arbitrary dimension.}
whose $\tau$ function and discriminant locus are
identical to those of $X_{n+1}$.  The set of genus-one fibered
Calabi--Yau manifolds with the same Jacobian fibration $J_{n+1}$ is
known as the Tate-Shafarevich group of $J_{n+1}$, denoted
$\sha(J_{n+1})$, and is identified with the discrete part of the gauge
group of F-theory following \cite{wilson-lines,triples}\footnote{The
  discrete part of a gauge group corresponds to the set of
  connected components of the group; a purely discrete gauge group
  is a finite group such as $\Z_n$}.  Note that $\sha(J_{n+1})$
represents not only a disjoint set of manifolds, but also includes an
abelian group structure \cite{MR0106226,MR0162806}.
Braun and Morrison
identified in the examples they studied an apparent deficit in the
number of scalar hypermultiplets required for gravitational anomaly
cancellation, when the massless scalars are identified only as the
complex structure moduli of the smooth genus one fibrations without
global section.  They resolved this apparent problem by identifying
additional massless hypermultiplets at nodes in the discriminant locus
of the Jacobian fibrations (more specifically, in the $I_1$ part of
that locus).  While this analysis supports the proposition that
genus-one fibrations without a global section are associated with
consistent F-theory backgrounds, it also raises several questions,
such as whether these backgrounds are connected to other F-theory
geometries or form a disjoint component of the moduli space of the
theory, and how the additional massless hypermultiplets should be
interpreted.  In this note, we show how these genus-one fibrations and
their Jacobians fit naturally into the connected moduli space of
Weierstrass models, and relate them to models with $U(1)$ gauge fields
arising from extra sections of the elliptic fibrations.  The structure
of $U(1)$ gauge fields in F-theory is rather subtle, as they are
determined by global features (the Mordell-Weil group) of an elliptic
fibration; F-theory models with one or more $U(1)$ fields have been the
subject of significant recent research activity (see for example
\cite{Park-WT,Park,Morrison-Park,Mayrhofer:2012zy,Braun:2013yti,Borchmann:2013jwa,Cvetic:2012xn,Borchmann:2013hta,Cvetic-Klevers-1,Cvetic-Klevers-2}).

In rough outline, the framework developed in this note is as follows:
over any complex $n$-dimensional base $B_n$, there is a space ${\cal
  W}$ of Weierstrass models, parameterized by the sections $f, g$ in
(\ref{eq:Weierstrass}).  Any Calabi-Yau $(n +1)$-fold with a genus-one
fibration has a multisection of some degree $k$, and its associated
 Jacobian fibration has a Weierstrass model which is generally singular
when $k>1$ (even in the absence of nonabelian gauge symmetry).
We can map the
 set ${\cal
  M}_k$ of genus-one fibrations with a $k$-fold multisection
(a ``$k$-section'', or when $k=2$, a ``bisection'') to a subset $
{\cal J}^k \subseteq {\cal W}$ of the set
of Weierstrass models, consisting of the Jacobians of those genus-one 
fibrations.  The set of elliptic fibrations with $k$
independent global sections (rank $r = k -1$ Mordell-Weil group) can also
be viewed through singular Weierstrass models as a subset ${\cal S}_k
\subseteq {\cal W}$ of the full space of elliptic fibrations.
For Calabi-Yau threefolds, these results follow for any $k$ from the result
of Nakayama \cite{Nakayama} and Grassi \cite{alg-geom/9305003}
that any elliptically fibered Calabi-Yau
threefold with section has a realization as a Weierstrass model that is
also Calabi-Yau;
as in the case of  the statements mentioned earlier concerning Jacobian
fibrations of Calabi-Yau genus-one fibrations,
this statement has not been mathematically proven  for
Calabi-Yau fourfolds, but there are no known examples to the contrary.
Furthermore, we have
\begin{equation}
 {\cal S}_k \subseteq {\cal J}^k \subseteq {\cal W}\,,
\label{eq:containment}
\end{equation}
meaning that the set of models with $k$ independent sections can be
viewed as a subset of the larger set of models with a $k$-fold
multisection.  
We give explicit formulae describing
these inclusions in the case $k =
2$ in the next section, but the inclusion ${\cal  S}_k \subseteq
{\cal  J}^{k}$ clearly holds for any $k$ since having $k$ independent
sections is a special case of having a $k$-fold multisection where the
$k$  sections can be given distinct global labels.
In particular, we can think of the multisection of an
$(n + 1)$-fold $X_{n + 1}\in {\cal J}^k$ as a branched cover of the
base; the multisection breaks into $k$ distinct global sections on a
subspace of moduli space where the branch points coalesce in such a
way as to give trivial monodromy among the branches.  In this picture,
going from a model in ${\cal S}_k$ to a model in ${\cal J}^k$ 
can be
interpreted physically as a partial Higgsing, where Higgsing some
charged matter fields breaks $U(1)^{k-1}$ to a discrete
subgroup, under which the remaining fields parameterizing ${\cal J}^k$
carry discrete charges.  In the case $k=2$, for example, we
can have matter fields with various integer-valued $U(1)$ charges;
if we Higgs matter fields with charge  $Q$, we break $U(1)$ 
to $\mathbb{Z}_Q$.

In the case $k=2$, we can also further analyze any model containing a
$U(1)$ by considering the explicit form of a Weierstrass model with
nonzero Mordell-Weil rank.  From this point of view we can demonstrate
that every $U(1)$ is associated geometrically with a nonabelian
$SU(2)$
(or larger) symmetry arising from a Kodaira type $I_2$ singularity along a
divisor on the base.  Starting with such an $SU(2)$ having both
adjoint and fundamental matter, there are several possible Higgsing
steps: the first leaves us with a $U(1)$ under which the remnant of
the adjoint matter has charge\footnote{A field is said to have ``charge $n$''
under a $U(1)$ gauge symmetry if it transforms as $e^{i n\theta}$ under a gauge
  transformation $e^{i \theta}\in U(1)$.} $2$ and the remnant of the
fundamental matter has charge $1$; the second Higgsing (of matter
fields of charge $2$) leaves us with gauge group $\mathbb{Z}_2$ under
which the remnant of the original fundamental matter is charged; a
final Higgsing of the fields originally carrying charge $1$ breaks the
residual discrete gauge group and moves the model out of ${\cal J}^k$
and into the moduli space ${\cal W}$ of generic Weierstrass models.

In \S\ref{sec:general} we describe the general framework for this
geometrical picture explicitly in the case $k = 2$, for a general base
manifold $B_n$.  In \S\ref{sec:example}, we show explicitly in 6D how
any $U(1)$ gauge field in an F-theory model can be associated with an
$SU(2)$ gauge group that has been Higgsed by an adjoint matter field,
and we look at several explicit examples.  \S\ref{sec:remarks}
contains some general remarks about the implications of this picture
for 6D and 4D supergravity theories, and some comments on further
directions for related research.

\section{General framework} 
\label{sec:general}

\subsection{Calabi-Yau manifolds with  bisections and with two different 
sections}
\label{sec:two-sections}

In \cite{Braun-Morrison}, an exercise in Galois theory provides an
equation for the Jacobian of a genus-one fibration with a bisection
\begin{equation}
y^2= x^3 - e_2 x^2 z^2 + (e_1 e_3 - 4e_0 e_4) x z^{4} 
-(e_1^2 e_4+ e_0e_3^2 -4e_0e_2 e_4) z^6 \,,
 \label{eq:Jacobian}
\end{equation}
where $e_0, \ldots, e_4$ are sections of various line bundles over the
base $B_n$ (to be determined below).  Completing the cube, changing
variables, and setting $z = 1$ puts this in Weierstrass form
\begin{equation}
 y^2 = x^3+ ( e_1e_3- \frac{1}{3} e_2^2 - 4e_0e_4) x
+(-
e_0e_3^2   +\frac{1}{3} e_1 e_2 e_3 -
\frac{2}{27} e_2^3 + \frac{8}{3} e_0e_2 e_4-e_1^2 e_4) \,. 
\label{eq:bisection}
\end{equation}
This parameterizes the set of all Jacobians of
genus-one fibrations over $B_n$ with
bisections, represented through the Weierstrass
models (of the Jacobian fibrations).  
In particular, this describes how ${\cal J}^2 \subseteq {\cal
  W}$ for any base $B_n$.

This class of Weierstrass models is closely related to the Weierstrass
form for elliptically fibered Calabi-Yau $(n+1)$-folds on $B_n$ with
two (different) sections.  Elliptically fibered Calabi-Yau manifolds with two
sections can be described as models with a non-Weierstrass
presentation (like the $E_7$ models of \cite{Klemm, Aluffi-Esole})
that are smooth for generic moduli.  All such $(n+1)$-folds, however,
also have a (possibly singular) description as Weierstrass models.  In
\cite{Morrison-Park}, the general form of such a Weierstrass model was
given as\footnote{We have modified eq.~(5.35) of \cite{Morrison-Park}
  by using a scaling $(f,g)\mapsto(i^4f,i^6g)$ to change the sign of
  $g$, and by changing $c_j$ in that paper to $e_j$ here
  ($j=0,1,2,3$).}
\begin{equation}
 y^2 = x^3+ (e_1e_3- \frac{1}{3}e_2^2 -b^2e_0) x
+ (-e_0e_3^2 +\frac{1}{3}e_1e_2e_3 - \frac{2}{27}e_2^3 + \frac{2}{3}b^2e_0e_2
 -\frac{1}{4}b^2e_1^2) \,.
\label{eq:two}
\end{equation}
Note that this equation is equivalent to (\ref{eq:bisection})
under the replacement 
$b^2 \rightarrow 4e_4$.  
The interpretation of this
analysis is that, as stated in the introduction,
\begin{equation}
 {\cal S}_2 \subseteq {\cal J}^2 \subseteq {\cal W} \,, 
\end{equation}
The condition $e_4 = b^2/4$ is precisely the condition that the
branching loci  of the bisection associated with a genus-one fibration  in
${\cal J}^k$ coalesce in pairs so that the total structure is that of
an elliptic fibration with two sections.  

In \cite{Braun-Morrison,Morrison-Park}, it was shown that for both an
elliptic fibration with two sections, and for a genus-one fibration
with a bisection, there is a natural model with a quartic equation
of the general form
\begin{equation}
 w^2= e_0u^{4} + e_1 u^3v + e_2 u^2 v^2 + e_3u v^3+ e_4 v^4\, .
\label{eq:quartic}
\end{equation}
If $e_4=b^2/4$, the equation can be rewritten
\begin{equation}
(w+\frac12 bv^2)(w-\frac12 bv^2) = u(e_0u^{3} + e_1 u^2v + e_2 u v^2 + e_3 v^3)\, ,
\label{eq:quartic-twosections}
\end{equation}
which makes the two sections manifest: they are given by $u=w\pm\frac12 bv^2=0$.
In general, when there are two sections one might need to make a linear
redefinition of the variables $u, v$ before \eqref{eq:quartic} can
be rewritten in the form \eqref{eq:quartic-twosections}, but after
such a linear redefinition it can always be done.

From the condition that $f, g$ in (\ref{eq:Weierstrass}) are sections
of the line bundles associated with $-4K, -6K$, we can characterize
the line bundles of which the $e_i$ and $ b$ are sections.  Focusing on
the $e_i$'s, we have
\begin{eqnarray}
-4K & = & 2[e_2]=[e_1]+[e_3]=[e_0]+[e_4]\,,\\
-6K & = &  2[e_1]+[e_4] =[e_0] +2[e_3] \,.
\end{eqnarray}
From $2[e_2]= -4K$, we have $[e_2]= -2K$.  We also note that $[e_0]=
-6K -2[e_3]$ must be an even divisor class. Choosing $[e_0]\equiv 2L$,
with $L$  the class of an arbitrary line bundle, we have
\begin{eqnarray}
\hspace*{0.1in}[e_0] & = & 2L \label{eq:e0}\\
\hspace*{0.1in}[e_1] & = &  - K + L\\
\hspace*{0.1in}[e_2] & = &  -  2K\\
\hspace*{0.1in}[e_3] & = &  - 3K - L\\
\hspace*{0.1in}[e_4] & = &  - 4K -2 L \label{eq:e4}\\
\hspace*{0.1in}[b] &=& -2K-L\, .
\end{eqnarray}
For any given base, $L$ can be chosen subject to the conditions that
$[e_1],[e_3]$ are effective divisors (if this condition is not
satisfied, then the only non-vanishing terms in the Weierstrass model
are those proportional to powers of $e_2$, and the discriminant
vanishes identically).  This constrains the range of
possibilities to a finite set of possible strata in the moduli space.
The consequences when $[e_4]$ and/or $[e_0]$ fail to be effective are
discussed in \S\ref{sec:enhancement}.

This analysis shows that for any Calabi-Yau manifold $X_{n + 1}$ that
is a genus-one fibration lacking a global section but having a
bisection, there is a Jacobian fibration $J_{n + 1}$, which has a
description as a Weierstrass model through (\ref{eq:bisection}).
Taking the limit $e_4\rightarrow b^2/4$ gives a Weierstrass model for an
elliptically fibered Calabi-Yau $(n + 1)$-fold with two sections,
which therefore has a Mordell-Weil group of nonzero rank.  In terms of
the  physical language of F-theory, as we describe in more detail in
the following sections,
this
corresponds to the reverse of a process in which a $U(1)$ gauge symmetry
is broken by matter fields of charge 2, leaving a discrete $\Z_2$
symmetry.  In \S\ref{sec:example} we describe several explicit
examples of this setup in 6D F-theory constructions.

\subsection{Singular fibers of type $I_2$ in codimension two}
\label{sec:I2}

One of the key features of the quartic models is the presence of singular
fibers in codimension two of Kodaira type $I_2$, observed in \cite{Morrison-Park}
in the $U(1)$ case, and in \cite{Braun-Morrison} in the bisection case.
When there is a $U(1)$, these singular fibers determine matter hypermultiplets
that are charged under the $U(1)$, and there can be different charges:
\cite{Morrison-Park}  focussed on the case when the charges are $1$ and $2$
only and found distinct geometrical interpretations for each of these.  
The geometric construction of $I_2$ fibers of charge $1$ under $U(1)$
extends to the case of a bisection (in the deformation from ${\cal S}_2$
to ${\cal J}^2$),
as we will now show explicitly.  As explained above, the corresponding
matter fields will be charged under the discrete $\mathbb{Z}_2$
gauge symmetry.
Both the bisection and $U(1)$ cases have a description in terms of the
quartic model (\ref{eq:quartic}).  We begin by considering the $I_2$
fibers in the genus one (bisection) case where $e_4$ is generic, and
then consider the limit where $e_4 = b^2$ is a perfect square,
corresponding to the $U(1)$ model.

\begin{figure}
\begin{center}
\includegraphics[scale=0.3]{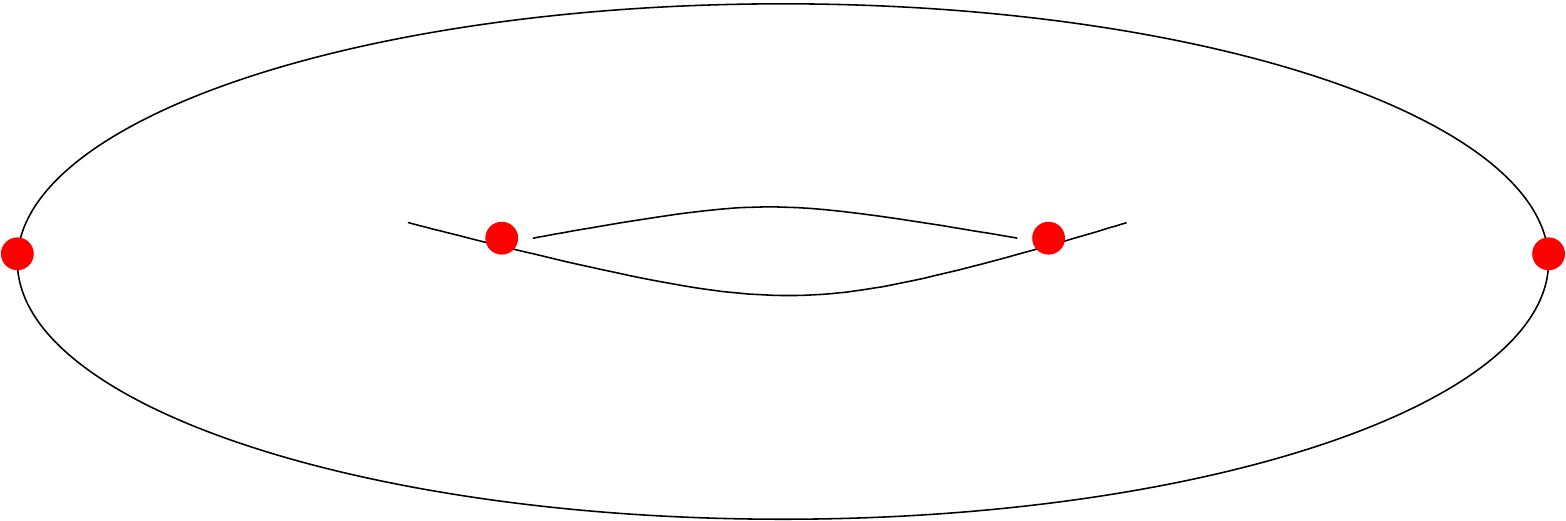} \qquad \qquad
\includegraphics[scale=0.3]{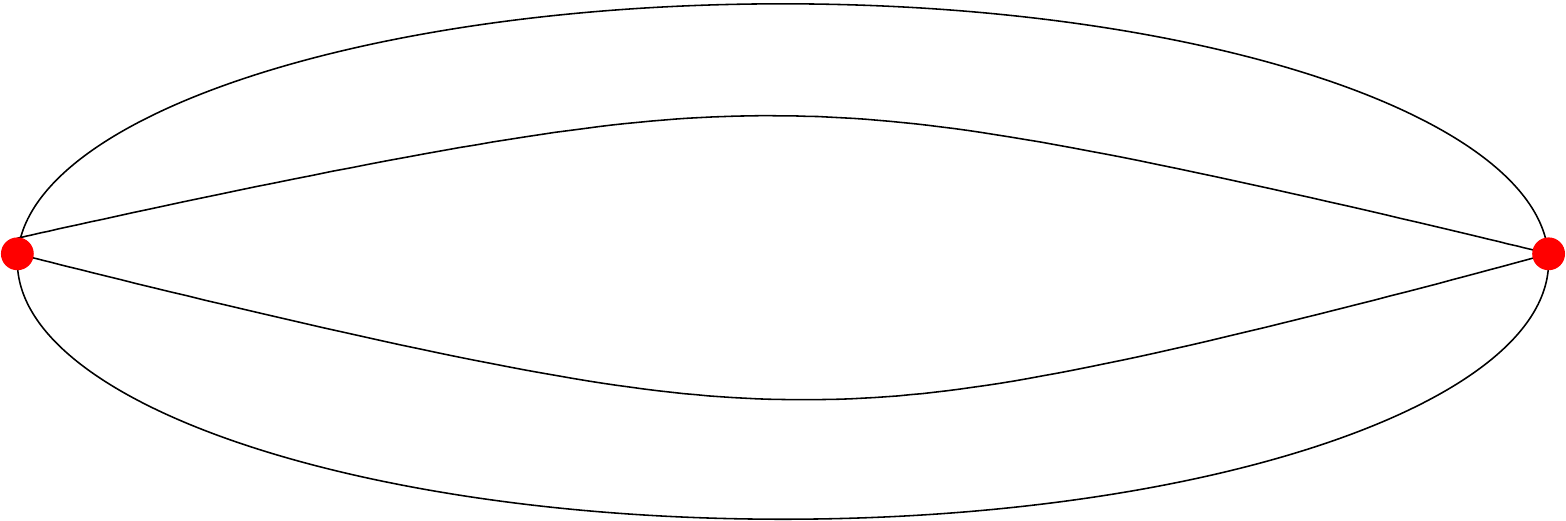}
\end{center}
\caption{Fiber of type $I_2$ as a degenerate branched cover}
\label{fig:I2}
\end{figure}

The curves of genus one in the quartic model are double covers of $\mathbb{P}^1$
branched in $4$ points, as illustrated in the left half of Figure~\ref{fig:I2}.
When the $4$ branch points come together in pairs, the resulting double
cover splits into two  curves of genus zero meeting in those two
double branch points, as illustrated in the right half of Figure~\ref{fig:I2}.
Such fibers in the family have type $I_2$ in the Kodaira classification.

Thus, to find such a fiber of type $I_2$ in the quartic model, we seek points
on the base $B_n$ for which the right-hand side of the
equation \eqref{eq:quartic} takes
the form of a perfect square.  As we explain 
in appendix~\ref{app:solving},
we can assume that
$e_4$ does not vanish at such points on the base (if the model is
sufficiently generic) and so we write our condition in the form
\begin{equation}
e_0u^{4} + e_1 u^3v + e_2 u^2 v^2 + e_3u v^3+ e_4 v^4
=e_4(\alpha u^2+\beta uv + v^2)^2 \,,
\label{eq:square}
\end{equation}
for some unknown $\alpha$ and $\beta$.  Multiplying out and equating
coefficients, it is easy to solve $\beta=e_3/2e_4$,
$\alpha=(4e_2e_4-e_3^2)/8e_4^2$ and then determine the remaining
conditions, which are:
\begin{align}
\label{eq:condition1}
e_3^4-8e_2e_3^2e_4+16e_2^2e_4^2-64e_0e_4^3 &= 0 \\ 
\label{eq:condition2}
e_3^3 - 4e_2e_3e_4 + 8 e_1e_4^2 &= 0 
\end{align}

To study the solutions of these equations, we introduce an auxiliary
variable $p$ and rewrite the equations as
\begin{align}
p^4-8e_2e_4p^2+16e_2^2e_4^2-64e_0e_4^3 &= 0 \\
p^3 - 4e_2e_4p + 8 e_1e_4^2 &= 0
\end{align}
In appendix~\ref{app:solving} we explain how to determine the condition 
\begin{equation}
(4e_0e_2-e_1^2)^2=64e_0^3e_4
\label{eq:condition}
\end{equation}
for these
equations to have a common root, and why that root is
\begin{equation}
p = \frac{8e_0e_1e_4}{4e_0e_2-e_1^2}
\label{eq:commonroot}
\end{equation}
when all of the coefficient functions $e_0$, \dots, $e_4$ are generic.
The points we seek can be described as solutions to \eqref{eq:condition}
which also satisfy $e_3=p$.

We now show how to count the solutions (i.e., the number of $I_2$ fibers
of this type), modifying an argument from \cite{Morrison-Park}.  Let us
take a limit, replacing $e_4$ with $\epsilon^2 e_4$ and then taking
$\epsilon$ very small (with both $e_0$ and $e_1$ of order $1$).  
Condition \eqref{eq:condition} then shows that
$4e_0e_2-e_1^2$ has order $\epsilon$, and \eqref{eq:commonroot}
shows that $p$ has order $\epsilon^2/\epsilon = \epsilon$.
It follows that any simultaneous solution to \eqref{eq:condition}
and $e_3=p$ can be deformed to a simultaneous solution to
\eqref{eq:condition} and $e_3=0$.  That is, the isolated $I_2$ fibers
are in one-to-one correspondence with the set
\begin{equation}
 \{ e_1^4-8e_0e_1^2e_2+16e_0^2e_2^2-64e_0^3e_4=0\} 
\cap \{ e_3=0\} \,.
\end{equation}
It follows that the number of $I_2$ fibers is
\begin{equation}
[4e_1] \cdot [e_3] = 4 (-K+L)\cdot (-3K-L) \,,
\end{equation}
since $e_1^4-8e_0e_1^2e_2+16e_0^2e_2^2-64e_0^3e_4$ is in class $[4e_1]$.

When $e_4=b^2/4$ so that we have a $U(1)$, 
the analysis above reproduces the count of $I_2$ fibers
found in \cite{Morrison-Park} which correspond to matter of charge $1$
under the $U(1)$ gauge group.  It was also observed there (and will
be mentioned again below) that when $U(1)$ is further enhanced to
$SU(2)$, this matter comes from matter in the fundamental representation
of $SU(2)$.  

On the other hand, the description of the matter of
charge $2$ in \cite{Morrison-Park} is a bit different:  it occurs
where $b$ and $e_3$ both vanish, and from 
\eqref{eq:quartic-twosections} and \eqref{eq:Jacobian}
we see that both the quartic
model and the Jacobian fibration have  conifold singularities
over each common zero of $b$ and $e_3$.
When we partially Higgs by relaxing the condition $e_4=b^2/4$, we
do a complex structure deformation of that conifold singularity,
giving a mass to the gauge field (as is standard in a conifold transition
\cite{GMS}).\footnote{The distinction between conifold singularities
which admit a K\"ahler small resolution and those which do not, and
the relation to massive gauge fields, has appeared a number of
times in the literature \cite{Jockers:2005zy,Buican:2006sn,
Grimm:2010ez,Grimm:2011tb, Braun:2014nva}.}  
It would be interesting to find a 
more geometric interpretation of this massive gauge field, perhaps
along the lines of \cite{GHS,DPS}.\footnote{We thank Volker Braun for
emphasizing the crucial role which must be played by massive gauge fields
in these models \cite{Volker-private}.}

The Weierstrass model of the Jacobian fibration also has a conifold singularity
corresponding to each $I_2$.  For models with two sections, these
conifold singularities have a (simultaneous) small resolution, as shown
explicitly in \cite{Morrison-Park} by blowing up the second section in
the Weierstrass model.  However, for Jacobians of models with a bisection,
the conifold singularities (i.e., the deformations of those 
singularities whose corresponding hypermultiplet
had charge $1$ before Higgsing) have no Calabi--Yau resolution, which
led to the question raised in \cite{Braun-Morrison} of whether
these are genuinely new F-theory models.

\subsection{Generalizations and geometry}
\label{sec:geometry}

In principle, our explicit analysis of bisections
could be extended to the spaces ${\cal
  J}^k$ of Jacobian fibrations associated with
genus-one fibered Calabi-Yau manifolds with $k$-sections and ${\cal
  S}_k$ of elliptically fibered Calabi-Yau manifolds with rank $r= k -1$ Mordell-Weil
group in a similar explicit fashion, at least for $k \leq 4$.
Explicit formulae for ${\cal S}_3,{\cal S}_4$, the generic forms of
elliptic fibrations with three and four sections respectively, were
worked out in \cite{Borchmann:2013jwa, Cvetic-Klevers-1} and \cite{Cvetic-Klevers-2}, and the
analogous formulae
for ${\cal J}^k$ are known \cite{AKMMMP} (although unwieldy to manipulate).
For $k = 3, 4$, the points in ${\cal S}_k$ correspond to singular
Weierstrass presentations of Calabi-Yau $(n+1)$-folds with 3, 4
independent sections, which have smooth descriptions similar to
the $E_6$ and
$D_4$ fibrations of \cite{Klemm, Aluffi-Esole}.

Even without an explicit description of the general form of a Jacobian
fibration with a $k$-section, it is clear that the  framework
described in the previous section should generalize.  In particular,
we expect that any Jacobian fibration $J_{n + 1} $
with a multisection will have a
discrete gauge group $\Gamma$ in the corresponding F-theory picture,
and that this will match the Tate-Shafarevich group $\Gamma =
\sha(J_{n+1})$.  There is a simple and natural geometric
interpretation of this structure in the M-theory picture.  When an
F-theory model on $J_{n + 1}$ is compactified on a circle $S^1$, it
gives a 5D supergravity theory that can also be described by a
compactification of M-theory on a Calabi-Yau $(n + 1)$-fold $Y_{n +
  1}$.  When there is a discrete gauge group $\Gamma$ in the 6D
F-theory model, a nontrivial gauge transformation (Wilson line) around
the complex direction gives a set of $| \Gamma |$ distinct 5D vacua
associated with $J_{n + 1}$.  In the M-theory picture this corresponds
precisely to the compactification on the set of distinct genus-one
fibered Calabi-Yau manifolds in the Tate-Shafarevich group
$\sha(J_{n+1})$.

\begin{figure}
\begin{center}
\includegraphics[scale=0.7]{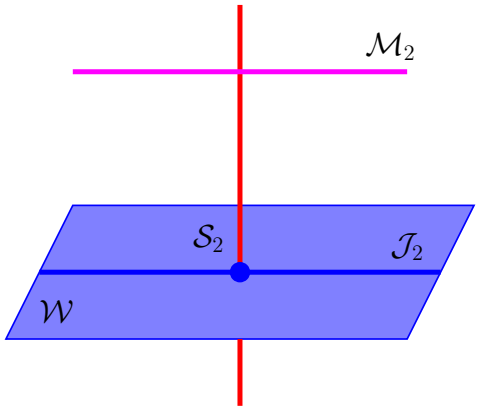}
\end{center}
\caption{Moduli spaces for M-theory compactifications on Calabi-Yau
  threefolds with different structures of sections (described in
  text).}
\label{fig:weierstrass}
\end{figure}

We can get a clear picture of the meaning of the multiple Calabi-Yau manifolds with
the same Jacobian fibration by considering the moduli space for the
compactified theory on a circle, which can be analyzed using M-theory.
We illustrate this in Figure~\ref{fig:weierstrass}, in which the moduli space 
${\cal W}$ of Weierstrass models (shown in blue) 
contains the subset ${\cal J}^2$
of Jacobians of models with a bisection, and this in turn contains
the subset ${\cal S}_2$ of Jacobians of models with two sections.
When there are two sections, the second Betti number of the Calabi-Yau
increases and there is an additional dimension in the K\"ahler moduli
space, which becomes  a modulus in the
compactified theory.  
We have illustrated this extra dimension as a red line emerging from the
${\cal S}_2$.

What initally seems puzzling is that while the Weierstrass
models of Jacobians of genus one fibrations with two sections
deform seamlessly to Jacobians of genus one fibrations with bisections
(by relaxing the condition that $e_4$ be a square), and similarly
the nonsingular fibrations with two sections deform seamless to genus
one fibrations with a bisection, the conifold singularities in the
Weierstrass model cannot be resolved in the bisection case.
The key to understanding this is to remember that the extra divisor
that is present when there are two sections (i.e., a $U(1)$)
allows an additional
K\"ahler degree of freedom which in particular allows us to specify 
the areas of the two components of an $I_2$ fiber independently.
On the other hand, when there is only a bisection, the homology classes
of those two components must each be one-half of the homology class
of a smooth genus-one fiber; thus, the two components must have
the same {\em area}.

The picture of the M-theory moduli space is thus completed by adding a
new component ${\cal M}_2$ of smooth
genus-one fibrations with a bisection, illustrated in purple in
Figure~\ref{fig:weierstrass}.  The new component must emerge from the
precise value of the additional K\"ahler classes (red line) at which
the two components of an $I_2$ fiber in the $U(1)$ case have an
identical area.  (Generally, the red line can be viewed as
parameterizing the difference of those areas.)  The additional
K\"ahler class in a $U(1)$ thus provides the connection between the
Weierstrass models (in which the area of one of the two components is
zero, corresponding to the conifold point without a Calabi-Yau
resolution) and the bisection models (in which the areas of the two 
components are equal).

Let us reiterate the crucial point:  away from the locus ${\cal S}_2$,
the complex structures on the Calabi-Yau manifolds represented by
the spaces ${\cal J}^2$ and ${\cal M}_2$ are different (and not even
birational to each other), and are only
related by the ``Jacobian fibration'' contruction.
However, they determine the same underlying $\tau$ function, so 
the F-theory models are identical.  Compactifying on a circle produces
two distinct geometries for M-theory models, which is precisely what one
expects for a discrete gauge symmetry.  Moreover, the ``extra'' hypermultiplets
have different but consistent explanations on the two components of
the M-theory moduli space.  Along ${\cal M}_2$, they are seen as geometric
$I_2$ fibers being wrapped by M2-branes, which were argued to have no
continuous gauge charges in \cite{Braun-Morrison} (although we now 
see that they carry $\mathbb{Z}_2$ gauge charges).  Along ${\cal J}^2$,
these same hypermultiplets are seen as complex structure moduli transverse 
to the ${\cal J}^2$ locus (moduli which are absent in ${\cal M}_2$).

One of the important lessons that we learn from this picture is that
it is important not to discard an F-theory model just because all of the
corresponding M-theory models after $S^1$-compactification are singular.
The lack of a nonsingular model means that the M-theory compactification cannot
be studied in the supergravity approximation without some additional
input to its structure, but such models must be included for a consistent
overall picture of the moduli spaces.

\subsection{Enhancement to $SU(2)$}
\label{sec:enhancement}

For any elliptically fibered threefold with nonzero Mordell-Weil rank,
we can carry the analysis of \S\ref{sec:two-sections} further, and
show that there is a limit in which an extra section in the
Mordell-Weil group transforms into a ``vertical'' divisor class lying
over a point in the base $B_n$.  In the F-theory language this
corresponds to an enhancement of the $U(1)$ gauge symmetry into a
nonabelian gauge group with an $\gsu_2$ 
gauge algebra (or in some special cases, a rank one enhancement of a
larger nonabelian gauge group).  At least at
the level of geometry, this shows that any $U(1)$ gauge group factor in
an F-theory construction can be found from the breaking of a
nonabelian group containing an $SU(2)$ subgroup by Higgsing a field in
the adjoint representation \cite{enhanced}.
This fits into a very simple and general
story associated with the Weierstrass form (\ref{eq:bisection}).
Examples of situations where $U(1)$'s can be ``unHiggsed'' in this
fashion were described in \cite{Morrison-Park, Grimm:2010ez}.
In most situations the unHiggsed model with a vertical divisor is
non-singular, though as we show explicitly in the following section,
in some cases a singularity is present which can be interpreted either
as a coupled superconformal theory, or as an indication that the
unHiggsed model is at infinite distance from the interior of moduli.

If the classes associated with the coefficients $e_0, e_4$ in
(\ref{eq:bisection}) are both effective, then all coefficients $e_0,
\ldots, e_4$ can generically be chosen to be nonzero, and we have a
family of Weierstrass models that characterize Jacobian fibrations
with a bisection, as discussed in \S\ref{sec:two-sections}.  Let us
consider what happens when $e_0$ and/or $e_4$ factorize or vanish
either by tuning or because the associated divisors are not effective
(in which case these coefficients would automatically vanish since
there would be no sections of the associated line bundles).

As described in  \S\ref{sec:two-sections}, if $e_4 = b^2/4$ is a perfect
square, then the bisection becomes a pair of global sections, and the
Mordell-Weil rank of the Jacobian fibration rises, which in the F-theory picture corresponds to
the appearance of a $U(1)$ gauge factor.  The equation is symmetric
under $e_i \rightarrow e_{4-i}$, however, so we can
also take $e_0 = a^2/4$ to produce a global section in another way.

In \cite{Morrison-Park} it was observed that the zeros of $b$ correspond
to the intersection points of the two sections, and that the further tuning
$b\to0$, which na\"\i{}vely would  place the two sections on top of each other,
in fact leads to a gauge symmetry enhancement to $SU(2)$.  We can
see this enhancement explicitly by choosing 
 $e_4 =b^2/4 = 0$ so that the
structure simplifies further, and the equation of the discriminant 
factorizes into the form 
\begin{equation}
\Delta := 4f^3 + 27g^2 = e_3^2 (-18 e_0 e_1 e_2e_3 + 4e_1^3 e_3-e_1^2
e_2^2 +4e_0 e_2^3 + 27e_0^2 e_3^2) \,.
 \label{eq:factored-discriminant}
\end{equation}
Since neither $f$ nor $g$ are generically divisible by $e_3$, this
corresponds to a family of singular fibers\footnote{Note that
when $I_2$ fibers occur in codimension one on the base, we associate them
to $\gsu_2$, but when they occur in codimension two on the base, they
are responsible for (charged) matter only and no additional gauge symmetry.}
 of Kodaira type $I_2$
along the divisor $\{e_3=0\}$, 
associated with an $\gsu_2$ 
Lie algebra component.  
(In some special cases the $\gsu_2$ can be part of a larger nonabelian
algebra, but we focus here on the generic $\gsu_2$ case for simplicity.)
In the F-theory picture the
transition from the model with $b = 0$ to the $U(1)$ model with $b \neq
0$ is described by the Higgsing of an $SU(2)$ gauge group by a matter
field in the adjoint representation.

Again, because the equation is symmetric, we can tune $e_0 = a^2/4 = 0$
in a similar fashion, giving a second $I_2$ singularity on the divisor
$\{e_1=0\}$.  In the F-theory picture this gives a second nonabelian gauge
group factor with an $\gsu_2$ algebra.  

This gives a very generic picture in which, when the divisor classes
$-K +L$ and $-3K-L$ are effective, we have a class of models with two
$A_1$ Kodaira singularities on the divisors $e_1, e_3$.  This
corresponds in the F-theory picture to a theory with gauge algebra
$\gsu_2 \oplus\gsu_2$.  When the divisor classes $L$ and $-4K-2L$ are
effective we can turn on terms $e_0 = a^2/4$ and/or $e_4 = b^2/4$ that
turn the ``vertical'' $A_1$ Kodaira singularities into global sections
(without changing $h^{1, 1} (X_{n + 1})$); this corresponds in
F-theory to Higgsing one or both of the nonabelian gauge groups
through an adjoint representation to the $U(1)$ Cartan generator.  When
$L, -4K-2L$ are nonzero classes, we can choose $e_0$ and/or $e_4$ to
be generically nonzero and non-square, which further breaks the $U(1)$'s
to a discrete $\Z_2$ symmetry.  Because the discrete $\Z_2$ symmetry
in the generic bisection model (\ref{eq:bisection}) is naturally
identified with the center of both $U(1)$ fields, we expect only one
$\Z_2$ in the center of the original nonabelian gauge group.  

This can
be seen geometrically by analyzing the charged matter
under each of the 
$SU(2)$ factors, following \cite{geom-gauge,anomalies}.  For any
F-theory model, the ``virtual'' or ``index''  spectrum 
of massless matter multiplets minus massless vector multiplets 
can be described in terms of
an algebraic cycle of codimension $2$ on the base $B_n$, to each component
of which is associated a representation of the gauge group.  
(For 6D compactifications, one then
just counts points in the $2$-cycle
to determine the multiplicity of the representation,
but for 4D compactifications there is an additional quantization 
which must be performed on each component of the $2$-cycle
to determine the multiplicity \cite{WitMF,Intriligator-jmmp},
which may depend on the G-flux; what
is fixed by the geometry is the set of representations which can appear
in the spectrum.)
For an $I_2$
fiber located along a divisor $\Sigma$, the virtual adjoint representation
in the matter spectrum is associated to the algebraic cycle
 $\Sigma\cdot \frac12(K+\Sigma)$, while the 
fundamental representation is associated to the cycle\footnote{More 
precisely, these cycles are determined by the intersections with
$\{f=0\}$ and $\{g=0\}$ in the Weierstrass model, as described
in \cite{anomalies}.}
$\Sigma\cdot
(-8K-2\Sigma)$.  Since we have bifundamental matter at the intersection
of $[e_1]$ and $[e_3]$, which counts as $2(-K+L)\cdot(-3K-L)$ fundamentals
for each of the $SU(2)$ factors, these bifundamentals account for
all of the fundamental matter.\footnote{Here we are using the fact that
when $\Sigma=-K+L$, we have $-8K-2\Sigma=2(-3K-L)$ and when $\Sigma'=-3K-L$,
we have $-8K-2\Sigma'=2(-K+L)$.}
Neither adjoints nor bifundamentals tranform nontrivially 
under the diagonal $\Z_2$ in the combined gauge group.  Since in the
M-theory picture the set of divisors must be dual to the set of curves
in the Calabi-Yau, the diagonal $\Z_2$ is not part of the gauge group
unless some field (associated with a curve in the resolved threefold)
transforms under it \cite{box-graphs,torsional}.  Thus, the full gauge group in the model with
$e_0 = e_4=0$ will be $(SU(2) \times SU(2))/\Z_2$ where the discrete
quotient is taken by the diagonal $\Z_2$.  Note that if either $[e_0]$
or $[e_4]$ is not effective then the corresponding $A_1$ (on $[e_1]$
or $[e_3]$) cannot be deformed away in the Weierstrass model while
preserving the element of $h^{1, 1} (X)$ in the form of a section; in
the F-theory picture this corresponds to an $SU(2)$ that does not have
massless matter in the adjoint representation.

We can confirm this analysis by exhibiting an explicit element of the
Mordell-Weil group of order 2, as predicted by \cite{pioneG} (see also
\cite{torsional}).  Namely, when $e_0=e_4=0$ the Weierstrass equation
\eqref{eq:bisection}
takes the form 
\begin{equation}
 y^2 = x^3+ (- \frac{1}{3} e_2^2 + e_1e_3) x
+ (\frac{1}{3} e_1 e_2 e_3 
  - \frac{2}{27} e_2^3) =(x+\frac13e_2)(x^2-\frac13e_2x-\frac29e_2^2+e_1e_3)\,.
\label{eq:su2su2}
\end{equation}
The factorization of the right side of the equation corresponds to a
point of order two on each elliptic fiber, and since this factorization
is uniform over the base, the locus $\{x=-\frac13e_2, y=0\}$ defines
a section which has order two in the Mordell-Weil group.
Note that either on the locus $e_1=0$ or on the locus $e_3=0$, the Weierstrass equation \eqref{eq:su2su2}
takes the form
\begin{equation}
 y^2 = x^3+ (- \frac{1}{3} e_2^2 + e_1e_3) x
+ (\frac{1}{3} e_1 e_2 e_3 
  - \frac{2}{27} e_2^3) =(x+\frac13e_2)^2(x-\frac23e_2)\,,
\label{eq:I2}
\end{equation}
showing that the $A_1$ singularity of the singular fiber is located at
$\{x=-\frac13e_2,y=0\}$ in each case, i.e., exactly at the section of order two.
This implies that the $\mathbb{Z}_2$ quotient is nontrivial on
each $SU(2)$, and thus that the global structure of the group must
be $(SU(2)\times SU(2))/\mathbb{Z}_2$ using the diagonal
$\mathbb{Z}_2$.

One additional complication that can arise in this picture is when
$-4K$ contains a divisor $A$ as an irreducible effective component.
In this case, there may be an automatic vanishing of $ f, g$ over $A$
giving a nonabelian gauge group, such as in 6D for the non-Higgsable
clusters of \cite{clusters}.  In this case, this component must be
subtracted out from $-4K$ in computing the complementary divisors on
which the $SU(2)$ factors reside, and some of the matter fields may
transform under the gauge group living on $A$ as well as one of the
$SU(2)$ factors.  We describe this mechanism further in the 6D context
in the following section.

The upshot of this analysis is that for any elliptically fibered
Calabi-Yau manifold with a nonzero rank Mordell-Weil group, a global
section can be associated with a divisor class $D =[e_3]$ to which the
section can be moved as an $A_1$
(or higher) Kodaira type singularity.  In the
language of F-theory geometry (without considering effects such as
G-flux relevant in four dimensions, \S\ref{sec:4D})
this means that any $U(1)$ gauge symmetry can be seen
as arising from a broken nonabelian symmetry.  Furthermore, there is an
intriguing structure in which for each such divisor class  $D$
there is a complementary divisor class
\begin{equation}
D' =[e_1] = -4K-D \,,
 \label{eq:complementary}
\end{equation}
that can (and in some cases must) also be tuned to support an $A_1$
singularity, which may be associated with a second independent
section.  In situations where $-4K$ has a base locus over which $f, g$
have enforced vanishing associated with Kodaira singularities giving
nontrivial gauge groups, the base locus must also be subtracted out in (\ref{eq:complementary}).
In the next section we give several explicit examples of how this
works in some 6D models.

\section{6D examples}
\label{sec:example}

The arguments given up to this point have been very general, and in
principle apply to elliptically fibered Calabi-Yau manifolds in all
dimensions where a suitable Weierstrass model is available.
In this section we consider some simple explicit examples of 6D
F-theory models to illustrate some of the general points.  
We begin by describing explicitly the way in which any $U(1)$ in 6D can be seen
as arising from an $SU(2)$ factor that has been Higgsed by turning on
a vacuum expectation value for an adjoint hypermultiplets.
We then describe some general aspects of models with bisections in
this context, and conclude by explicitly analyzing various possible
ways in which the ``unHiggsing'' to $SU(2)$ may encounter problems
with singularities.  While such singularities do not arise in most
cases, we identify one situation where such a singularity arises,
which can only be removed by blowing up the F-theory base manifold.

Before beginning, let us recall how the various moduli spaces of
Weierstrass models are linked together through transitions involving
coupling to 6D superconformal field theories \cite{Seiberg:1996qx},
sometimes called ``tensionless string
transitions''  \cite{Ganor-Hanany,Seiberg-Witten}.  
As we tune the coefficients of a Weierstrass model
over a fixed base $B_2$, various singularities are encountered that
have explanations in terms of nonabelian gauge symmetry or the massless
matter spectrum.  However, if a singular point $P$ is encountered at which
$f$ has multiplicity at least $4$, and $g$ has multiplicity at least
$6$, the model has a superconformal field theory sector 
and another branch emerges in
which a tensor multiplet is activated \cite{Morrison-Vafa-II,Heckman:2013pva}.
The other branch consists of
Weierstrass models over the blowup $\operatorname{Bl}_P(B_2)$ of $B_2$
at $P$, and the area of the exceptional curve of the blowup serves
as the expectation value of the scalar in the new tensor multiplet.
We generally refer to such points as ``$(4,6)$ points.''

Even more special is the case in which either $(f,g)$ have
multiplicities at least $(8,12)$ at a point, or have  multiplicities of at
least $(4,6)$ along a curve.  In this case, the total space of the
fibration is not Calabi-Yau, and in fact any resolution of the space
in algebraic geometry has no nonzero holomorphic $3$-forms.  It is
known that the points in the moduli space of Weierstrass models at
which such singularities occur are boundary points of moduli at
infinite distance from the interior of the moduli space
\cite{hayakawa,wang}.

\subsection{$U(1)$  from a Higgsed $SU(2)$ in 6D}
\label{sec:Higgsing-proof}

In six dimensions, we can demonstrate explicitly that in most
situations a $U(1)$ can be enhanced to an $SU(2)$ in a conventional
F-theory model on the same base (i.e., one not involving a superconformal
theory or at infinite distance from the interior of the moduli space) 
by considering general classes of
acceptable $U(1)$ model in which tuning $b^2 \rightarrow 0$ in
(\ref{eq:two}) need not introduce a $(4, 6)$ point.  We can also
identify some situations in which this tuning does necessarily lead to
such a singularity.  A forced $(4,6)$ point can in principle
occur in one of two ways: first, if $[e_3]$ contains a curve of
negative self intersection over which $f, g$ are required to vanish to
high degree, and second if $[e_3]$ has nonzero intersection with
another curve $C$
or combination of curves $A, B, \ldots$ over which $f, g$ vanish to high enough degree to
force a $(4, 6)$ vanishing at an intersection point.  We outline the
general structure of the analysis here, and describe some special
cases in the later parts of this section.

First, let us consider the case where $C$ is a curve in the class
$[e_3]$, $C$ is irreducible, and $[e_1] = -4K-[e_3]$ does not contain
as irreducible components any curves of self-intersection below $-2$.
We consider the Weierstrass model of the form (\ref{eq:two}), and take
the limit as $b^2 \rightarrow 0$, which produces an $SU(2)$ over $C$
with matter in the adjoint representation.  For the enhancement to
$SU(2)$ with an adjoint or higher representation
to occur on a curve $C$, the curve must have
genus $g > 0$.  
This follows from the general result \cite{KPT} that every
representation of $SU(N)$ with a Young diagram having more than one
column makes a positive contribution  to the genus of the curve
through the anomaly equations.
It was shown in \cite{clusters} that a curve of
  positive genus cannot have negative self-intersection without
  forcing a $(4, 6)$ vanishing all along the curve.  So $C \cdot C \geq 0$ and
  $f, g$ cannot be required to vanish on the irreducible curve for a generic
  Weierstrass model over the given base.  To see where there are
  enhanced singularities at points on $C$ in the $SU(2)$ model, we can use
  the 6D anomaly cancellation conditions \cite{gswest, Sagnotti,
    Erler, Sadov, Grassi-Morrison, KMT-II}.  For a generic curve $C$
  in the class $[e_3]$, where $SU(2)$ matter is only in $A$
  hypermultiplets that transform under the adjoint (symmetric)
  representation and $x$ fundamental hypermultiplets, the anomaly
  conditions read
\begin{eqnarray}
K \cdot C & = &  \frac{1}{6} \left[ 4 (1-A)-x \right] \\
C \cdot C & = &  -\frac{1}{3}  \left[ 8 (1-A) -x/2 \right] \,.
\end{eqnarray}
Solving these equations gives
\begin{equation}
A = g = 1 +\frac{1}{2} ( K \cdot C + C \cdot C)
\label{eq:A}
\end{equation}
and
\begin{equation}
x =2 C \cdot (-4K-C) =2 [e_3] \cdot[e_1] \,.
\label{eq:x}
\end{equation}
When $e_0 = a^2/4 = 0$ and there are $SU(2)$ gauge factors supported on
both $e_1$ and $e_3$, this shows that all matter fields -- and hence
all enhanced singularities -- arise at the intersection points between
these two curves.  
Note that this is the same conclusion about the matter spectrum that
we reached in \S\ref{sec:enhancement} in a different way.
Note also that the location of the singularities associated with the
matter charged under the $SU(2)$ on $C$ is the same whether or not we
take the $a \rightarrow 0$ limit, and that this matter corresponds to
the extra charged matter fields found on the $I_2$ locus in
\S\ref{sec:I2}.

Additional complications can arise if $e_1$ or $e_3$ are reducible,
particularly when either or both contain irreducible factors that
carry nontrivial Kodaira singularities.  In such cases, $f, g$ will
vanish on the associated curve $A$, with an extra nonabelian gauge
group factor, according to the classification of non-Higgsable
clusters in \cite{clusters}.  
To show when a $U(1)$ that arises in a Weierstrass form
(\ref{eq:two}) can be associated with a broken $SU(2)$ 
in a conventional F-theory model without changing the base,
we need to
prove that in these cases a $(4, 6)$ point cannot be introduced
by taking the $b \rightarrow 0$ limit.
There can also be more complicated singularities introduced if the
curve $C$ is not a generic curve in the class $[e_3]$ and  itself has
singularities.  There is not yet a complete dictionary relating
codimension two singularities of this type to matter representations,
though there has been some recent progress in this direction
\cite{singularities, Esole-Yau, Grassi:2013kha}.  We do not consider such
cases here in any detail, though an example is discussed in
\S\ref{sec:6D}; here we assume that the curve $C$ is taken to be
generic in the class $[e_3]$, so the statement that a $U(1)$ can be
viewed as a Higgsing of an $SU(2)$ model  should be understood as
involving the Higgsing of an $SU(2)$ model with a generic $C$ given
$[e_3]$, with further tuning of $C$ carried out as necessary to
achieve the given $U(1)$ model of the form (\ref{eq:two}).

If $[e_3]$ intersects  a curve
$A$ in $[e_1]$ that carries a nonabelian gauge group $G_A$
(again, assuming $A$  is a generic curve in its class), some of
the matter charged under the $SU(2)$ living on the curve $C$ will also
be charged under $G_A$.
This must occur in such
a way that setting $b^2 \rightarrow 0$
does not increase the degree of
vanishing of $f, g$ on $A$, or the spectrum of fields charged under
the $U(1)$ would not match the spectrum of fields charged under the
$SU(2)$ in the $b \rightarrow 0$ limit determined as above by the
anomaly conditions.  Indeed, explicit analysis of the possibilities
shows that  such an intersection can occur only when $A$ is a
$- 3$ or $-4$ curve.  In these cases, when $[e_3] \cdot A \neq 0$, 
the degrees of vanishing of $f, g$ on $A$ are increased above the
minimal Kodaira levels, and $G_A$ carries an enhanced gauge group with
charged matter that also carries charges under the $U(1)$ or $SU(2)$
on $C$ in a consistent fashion.  When $A$ is a $-5$ (or less)
curve, there is a
$(4, 6)$ point on $A$ even in the $U(1)$ model (\ref{eq:two}),
so no such conventional $U(1)$ theory can be constructed.
We consider some explicit examples of these cases in the subsequent
sections and demonstrate the unconventional presence of a
superconformal theory  explicitly for $-5$ curves
in \S\ref{sec:f5}.

In a similar fashion, we can analyze the special cases where $e_3$
contains a curve $D$ of negative self-intersection as an irreducible
component.  Note that if $d | e_3$ and also $d | b$, then we can move
the factor of $d$ from $e_3$ into $e_1$ (with two factors of $d$
extracted from $b^2$ and moved into $e_0$).  Thus, if $[e_3]$ contains
$[d]$ as a component, and $b = -2K-L$ is such that $[b]-[d]$ is
effective, we can tune $b = db'$ and the analysis becomes that of the
previous case.  So we need only consider situations where $e_3$
contains an irreducible component $D$ that is not a component of $b$.
It turns out this is possible for curves of self-intersection $-3, -4,
-5,$ and $-6$; in each of these cases there are configurations where
$e_3$ contains such curves as a component but $b$ does not.  When the
parameter $b$ is tuned to vanish, the enhancement to $SU(2)$ is
combined with an enhancement of the gauge group over $D$ in a way that
is consistent with anomaly cancellation and does not introduce $(4,
6)$ points.  For curves of self-intersection $-7$ or below, the
$U(1)$ model already has $(4, 6)$ singularities, so there are no
conventional models.  
We give some examples of these kinds of
configurations in the subsequent parts of this section.

Although a single curve of negative self-intersection contained in
$e_3$ does not lead to a problematic singularity, there are also
situations where $e_3$ contains a more complicated configuration of
intersecting negative self-intersection curves.  In particular, there
exist non-Higgsable clusters identified in \cite{clusters} that
contain intersecting $-3$ and $-2$ curves.
In such a situation, as we show explicitly below, a $(4, 6)$
point can arise at the intersection between these curves when a
$U(1)$ is unHiggsed  to $SU(2)$ by taking the $b \rightarrow 0$
limit.  This is the one situation we have clearly identified in which
such a singularity can arise.

This argument shows that a $U(1)$ gauge factor in a 6D F-theory model
over any base can be viewed as arising from an $SU(2)$ gauge group
supported on a corresponding effective irreducible divisor class
$[e_3]$, after Higgsing a matter hypermultiplet in the adjoint
representation; in a wide range of situations the unHiggsing results in a conventional
F-theory model with reduced Mordell-Weil rank, though in certain
special cases the unHiggsing either gives rise to a model which is
coupled to a superconformal theory or is at infinite distance from
the interior of moduli space. This 
general framework
gives
strong restrictions on the ways in which $U(1)$ factors can arise in
6D F-theory models, and illuminates the structure of the Mordell-Weil
group for elliptically fibered Calabi-Yau threefolds over general
bases.

\subsection{6D theories on $\P^2$ with two sections or a bisection}
\label{sec:p2}

As a simple specific example of a class of 6D theories that illustrate
the general structure of models with bisection, two sections ($U(1)$)
and enhanced $(SU(2) \times SU(2))/\Z_2$ gauge group, we consider the
case of 6D F-theory compactifications on the simplest base surface
$B_2 = \P^2$.  Models of this type with $U(1)$ fields were considered
from the point of view of supergravity and anomaly equations in
\cite{Park-WT}, and an explicit F-theory analysis and Calabi-Yau
constructions were given in \cite{Morrison-Park}.  In this case, $- K
= 3H$, where $H$ is the hyperplane (line) divisor with $H \cdot H =
1$.  Tuning an $I_2$ singularity along a degree $d$ curve $C$ in $\P^2$
by adjusting  the degrees of vanishing of $f, g, \Delta$ along $C$ to be
$0, 0, 2$, respectively,  gives an
F-theory model with gauge group $SU(2)$.  A generic curve of degree
$d$ has genus $g = (d-1) (d-2)/2$, and the associated $SU(2)$ gauge
group has a matter content consisting of $g$ massless hypermultiplets
in the adjoint representation and $24 d-2d^2$ multiplets in the
fundamental representation (note that for $SU(2)$, unlike $SU(N)$ for
$N> 2$, the antisymmetric representation is trivial).  By tuning higher order singularities in
the curve $C$, some of the adjoint matter fields can be transformed
into higher-dimensional matter fields, with a simple relation between
the matter representations and contribution to the arithmetic genus of
$C$, as described in \cite{KPT, singularities}.

To describe the class of Calabi-Yau threefolds on $\P^2$ associated
with a Jacobian fibration with a bisection, we consider Weierstrass
equations of the form (\ref{eq:bisection}), where the classes of the
$e_i$ are given in (\ref{eq:e0}--\ref{eq:e4}).  We parameterize the
set of models of interest by $[e_3] =-3K-L = mH$, where $m$
corresponds to the degree of a curve in the class $[e_3]$.  For any
$m$ in the range $0 \leq m \leq 12$ there is a class of Weierstrass
models of the form (\ref{eq:bisection}) that give ``good'' F-theory models
without $(4, 6)$ points (points which, if present,  would 
involve coupling to superconformal field theories
or would violate the Calabi-Yau
condition).  The generic model in each of these classes corresponds to
a Jacobian fibration, and in the F-theory picture there is a discrete
$\Z_2$ gauge group, with a number of charged matter hypermultiplets.
For $3 \leq m \leq 9$, both $[e_0]$ and $[e_4]$ are effective; in this
range of models, there is a subset of models with $e_4= b^2/4$ a perfect
square, giving an extra section contributing to the Mordell-Weil rank,
which is associated in the F-theory picture with a $U(1)$ gauge
factor, and there is also a (partially overlapping)
subset of models with $e_0 = a^2/4$ with
another $U(1)$ factor.  Either or both of these $U(1)$ factors can be
further enhanced to an $SU(2)$ by fixing $b^2 = 0$ or $a^2 = 0$.  When
both factors are enhanced ($e_0 = e_4 = 0$) the total gauge group is
$(SU(2) \times SU(2))/\Z_2$.  When $m < 3$ or $m > 9$ the story is
similar but one of the two $SU(2)$ factors is automatically imposed by
the non effectiveness of the divisor $[e_4]$ or $[e_0]$; in these
cases there is only one possible $U(1)$ factor.

This class of models can be understood most easily in the F-theory
picture starting from the locus $e_0 = e_4 = 0$ where the gauge
algebra is $\gsu_2\oplus\gsu_2$.  In this case, the two $\gsu_2$ summands
are associated with 7-branes wrapped on divisors $D, D'$
given by curves of degrees $m$ and $12-m$ in the classes
$[e_3],[e_1]$.  The spectrum of the theory consists of $m (12-m)$
bifundamental hypermultiplets (associated with the intersection points
of $D, D'$), and $(m-1) (m-2)/2, (11-m) (10-m)/2$ fields in the
adjoint representation of each $SU(2)$.  The limiting cases $m = 0,
12$ correspond to situations with only a single $SU(2)$ factor and no
fundamental hypermultiplets.  In all cases, an $SU(2)$ on a curve of
degree $d \geq 3$ has adjoint hypermultiplets, of which one can be
used to Higgs the nonabelian gauge group to a $U(1)$.  Under this
Higgsing, the remaining adjoints become scalar fields of charge 2
under the resulting $U(1)$, while fundamentals acquire a charge of 1.
When $3 \leq m \leq 9$, such Higgsing to abelian factors is possible
for both $SU(2)$ factors; for other values only one of the groups can be
Higgsed.  Once one or both of the nonabelian factors are Higgsed to
$U(1)$ fields, a further breaking can be done by making $e_0$ or $e_4$ a
generic non-square.  This corresponds to using the charge 2 fields to
Higgs the $U(1)$ to a discrete gauge group $\Z_2$.  Under this Higgsing,
the charge 1 fields retain a charge under the discrete gauge group.
It is straightforward to check that the numbers of fields in each of
these models satisfies the gravitational anomaly cancellation
condition $H-V = 273-29T$, and matches with the results of
\cite{Park-WT, Morrison-Park, Braun-Morrison} for the various
component theories.
In particular, note that for $m = 3$ the $SU(2)$ gauge group on $D
=[e_3]$ only has a single adjoint field, so after breaking to $U(1)$
there are only charge 1 hypermultiplets.  Thus, in this case there is
no way of breaking to a model with a bisection and residual discrete
gauge group.  Note also that by tuning a non-generic singularity on
the curve $C$ carrying an $SU(2)$ factor, it should be possible to construct
higher dimensional representations of $SU(2)$, which will correspond to
larger charges $Q \geq 3$ after breaking to $U(1)$, and which can give
rise to higher order discrete gauge groups $\Z_Q$.  We return to this
issue in \S\ref{sec:6D}.
In Table~\ref{t:charges}, we provide an explicit list of the charges
that arise for the $SU(2)$ and $U(1)$ factors in the various
relevant components of the Weierstrass moduli space ${\cal S}_2,{\cal J}^2$

\begin{table}
\begin{center}
\begin{tabular}{| c | rr |rr |}
\hline
$m$ & $n_{\rm a}$ & $n_{\rm f}$ &
$n_2$ &  $n_1$\\
\hline
1 & 0 & 22 &--&--\\
2 & 0 & 40 &--&--\\
3 & 1 & 54 & 0 & 108\\
4 & 3 & 64 & 4 & 128\\
5 & 6 & 70 & 10 & 140\\
6 & 10 & 72 & 18 & 144\\
7 & 15 & 70 & 28 & 140\\
8 & 21 & 64 & 40 & 128\\
9 & 28 & 54 & 54 & 108\\
10 & 36 & 40 & 70 & 80\\
11 & 45 & 22 & 88 & 44\\
12 & 55 & 0 & 108 & 0\\
\hline
\end{tabular}
\end{center}
\caption[x]{\footnotesize  
Table of $SU(2)$ charges in adjoint and fundamental, and $U(1)$ charges in
associated theory,
when $e_3$ describes a curve of degree $m$ in $\P^2$.
Note that $SU(2)$ and $U(1)$ charges match with Higgsing description
($n_2 = 2 (n_{\rm a} -1), n_1 = 2n_{\rm  f}$) as well as with
charges computed in \cite{Park-WT, Morrison-Park}.
Note also that $n_1$ matches for $m, 12-m$, in agreement with the general
picture that all charged matter lies on intersection points of
$[e_1],[e_3]$ for the $(SU(2) \times SU(2))/\Z_2$ theory.
}
\label{t:charges}
\end{table}

\subsection{6D theories on $\F_0 =\P^1 \times\P^1$}

A similar structure will hold on any base $B_2$ that supports an
elliptically fibered Calabi-Yau threefold; a classification of such
bases was given in \cite{clusters}, and a complete list of toric bases
was given in \cite{toric}.  As another example we consider the
Hirzebruch surface $\F_0 =\P^1 \times\P^1$.

For $\F_0$, a basis of $h^{1, 1}$ is given by $S, F$ with $S \cdot S =
F \cdot F = 0, S \cdot F = 1$.  A divisor $D = aS + bF$ is effective
if $a, b \geq 0$, and the anticanonical class is $-K = 2S + 2F$.
The genus of a curve in the class $C = aS + bF$ can be computed 
as
\begin{equation}
(K + C) \cdot C = 2g-2  = 2(ab-a-b) \,.
\end{equation}
The genus is nonzero iff $2 \leq a, b$.

The range of possible models (\ref{eq:bisection}) with a bisection is
thus given by $[e_3] = aS + bF$ with $0 \leq a, b \leq 8$.
The values of $a, b$ for which the curves $[e_3], [e_1]$ both have
nonzero genus and associated $SU(2)$s can be broken is $2 \leq a, b \leq
6$.  Within this range we have the full set of possible enhancements
of a model of type (\ref{eq:bisection}); there is a model with $(SU(2)
\times SU(2))/\Z_2$ symmetry, where either or both $SU(2)$'s can be
broken to $U(1)$ or further to the discrete $\Z_2$ symmetry.  Again,
counting charged multiplets confirms that anomaly cancellation in both
the nonabelian and abelian theories matches with the Higgsing
process. 
The spectrum of charged matter fields for an $SU(2)$ tuned on a divisor
$aS + bF$ consists of $g = ab-a-b + 1$ adjoints and $16 (a + b)-4ab$
fundamentals.  As in the $\P^2$ case, the number of fundamental fields
is symmetric under $a \leftrightarrow 8-a, b \leftrightarrow 8-b$
($e_1 \leftrightarrow e_3$), corresponding to the fact that all
charged matter in the overall $(SU(2) \times SU(2))/\Z_2$ theory is
contained in the adjoints and $8 (a + b)-2ab$ bifundamental fields.

\subsection{6D theories on $\F_3$}
\label{sec:f3}

Some interesting points are illuminated by examples on the Hirzebruch
surface $\F_3$.  Here we have a basis of curves $S, F$ with $S\cdot S=
-3, S\cdot F = 1, F \cdot F = 0$.  The canonical class is $-K = 2S+
5F$, and there is an automatic vanishing of $f, g, \Delta$ to degrees
$2, 2, 4$ giving an $SU(3)$ gauge group supported on the divisor $S$
in a generic elliptic fibration.

The simplest irreducible
curve $e_3$ that can give rise to a $U(1)$ factor is $C = 2\tilde{S} =
2S+ 6F$, since $e_4$
must be effective; a generic curve in this class $C$ is irreducible and has genus 2.  Choosing
\begin{equation}
[e_3] = 2S+ 6F, \; \; \Rightarrow \; \;
[e_1] = 6S+ 14F \,.
\label{eq:choice}
\end{equation}
We note that $[e_1] \cdot S= -4$, so $[e_1]$ contains $S$ as
an irreducible component with multiplicity at least $2$.  There is an $SU(3)$ over $S$, but this does not
cause any problems since $S\cdot[e_3] = 0$ so there is no matter
charged under the $SU(3)$ that interacts with the $SU(2)$ supported on $C$
or the corresponding $U(1)$ when $e_4 = b^2/4 \neq 0$ or discrete group
$\Z/2$ when $e_4$ is non-square.  So this case works like the others
above,
with $[e_1]  \cdot[e_3]= 28$ bifundamental matter fields.

The next case of interest is 
\begin{equation}
[e_3] = 2S+ 7F, \; \; \Rightarrow \; \;
[e_1] = 6S+ 13F \,.
\label{eq:choice2}
\end{equation}
In this case the curve $C$ defined by the vanishing locus of $e_3$ is
generically a smooth irreducible curve of genus 3.  In this case, $C
\cdot S= 1$, so there is matter charged under the gauge group lying
over $S$.  To analyze this explicitly, we see that $[e_0] = 8S+
16F,[e_1]= 6S + 13F,[e_2] = 4S+ 10F$ contain the irreducible component
$S$ with multiplicities 3, 2, and 1 respectively.  From
(\ref{eq:two}), (\ref{eq:factored-discriminant}), this shows that $f,
g, \Delta$ vanish to degrees $2, 3, 6$ at generic points over $S$, and
to degrees $2, 3, 8$ at points of intersection $[e_3] \cdot S$.  As
discussed in \S\ref{sec:Higgsing-proof}, in this situation the gauge
group over the $-3$ curve has an algebra that is larger than the
minimal $\gsu_3$ for a generic model over $\F_3$.  The configuration
in this case is similar to the
$(-3, -2)$ non-Higgsable cluster \cite{clusters}, in which a $-3$
curve carries a
$\gg_2$ algebra, and there are  matter fields charged under both this
algebra and an $\gsu_2$ on a curve that intersects the $-3$ curve. 

We can also analyze the case where $C = - K =2 S + 5F$, where $C$ is
reducible and contains $S$ as a component in a similar fashion, which
also gives a $(2, 3, 6)$ vanishing on $S$, with a similar interpretation

\subsection{$\F_4$}
\label{sec:f4}

The analysis in the case of a $-4$ curve in $\F_4$ is similar to $\F_3$.
The curve $S$ has $S\cdot S= -4$.  For the minimal irreducible case $[e_3] = 2S+ 8F$, there is an
$SU(2)$ with adjoint matter that does not intersect $S$.  For $[e_3] =
2S+ 9F$, we have
\begin{eqnarray}
\;[e_2] & = &  4S+ 12F = S+ X^{(2)}_{\rm eff}\\
\;[e_1] & = &  6S+ 16F =2 S+ X^{(1)}_{\rm eff}\\
\;[e_0] & = &  8S+ 20F = 3S+ X^{(0)}_{\rm eff}
\end{eqnarray}
where $X^{(a)}_{\rm eff}$ are effective divisors that contain no
further components of $S$.  We can read off the order of vanishing of
$f, g, \Delta$ from (\ref{eq:two}) and
(\ref{eq:factored-discriminant}) as $(2, 3, 6)$ on $S$, enhanced
to $(2, 3, 8)$ on $[e_3] \cdot S$, so again we have hypermultiplets charged
under the gauge group on $S$ as well as the $SU(2)$ on $[e_3]$. 
For curves such as $[e_3] = - K = 2 S + 6F,$ where $[e_3]$ contains
$S$ as an irreducible component, a similar analysis holds.

\subsection{$\F_5$ and $-5$ curves}
\label{sec:f5}

Now let us consider a $-5$ curve, beginning with the case of $\F_5$.  As
in the previous cases, for $[e_3] = 2S+ 10F$, there is no intersection
with $S$ and the $SU(2)$ story is as above.  For the next interesting
case, however, we have
\begin{eqnarray}
\;[e_3] & = &  2S+ 11F\\
\;[e_2] & = &  4S+ 14F =  2S+ X^{(2)}_{\rm eff}\\
\;[e_1] & = &  6S+ 17F = 3 S+ X^{(1)}_{\rm eff}\\
\;[e_0] & = &  8S+ 21F = 4S+ X^{(0)}_{\rm eff} \,.
\end{eqnarray}
Now, analyzing (\ref{eq:two}) and (\ref{eq:factored-discriminant}) we
find vanishing orders of $f, g, \Delta$ on $S$ of $(3,  4, 9)$,
enhanced to $(4, 6, 12)$ on $S\cdot[e_3]$, even when $b^2 \neq 0$.
Thus, there cannot be a $U(1)$ model based on (\ref{eq:two}) using $e_3
= 2S+ 11F$ (unless the intersection point is blown up, giving a model
on a different base).

More generally, we can show that a $U(1)$ based on an extra section
can never be constructed on any curve $[e_3] = C$ if $C \cdot A > 0$
  for some curve $A$ of self-intersection $-5$ or less.  The argument
  basically follows exactly the same steps as above.  In general, as
  described in \cite{clusters}, from $[e_2] = -2K$ it follows that
  $e_2$ vanishes to degree 2 on $A$ just as in the $\F_5$ case.  We
  have $-4K \cdot A = - 12$ and $[e_3] \cdot A > 0$, so $[e_1] \cdot A
    = (-4K-[e_3]) \cdot A < -12$ and $e_1$ vanishes to degree 3 on
      $A$.  From $[e_3] \cdot A > 0,$ it follows that $L \cdot A \leq
    -10$, so $[e_0] \cdot A \leq -20$, and $e_0$ vanishes to order 4
    on $S$.  Thus, no $U(1)$ can be built using (\ref{eq:two}) on any
    curve $e_3$ that has positive intersection with a curve $A$ of
    self-intersection $-5$.  The condition on each term is stronger as
    the self-intersection decreases further, so the same result holds
    for any curve of self-intersection ${}< -5$.

Now,  let us consider the case that $e_3$ itself has a $-5$  curve $D$ as a
component.  For this to happen we must have $[e_3]\cdot D < 0$, but as
  argued in \S\ref{sec:Higgsing-proof} we should also have $[e_4]
  \cdot D \geq$ 0, or we could move the associated factor out of $e_3$
  and into $e_1$.  This can lead to a conventional model when $[e_3]
  \cdot D =
  -2$ or $-3$.  In these cases, $e_0, e_1, e_2, e_3$ vanish to degrees
  $3, 2, 2, 1$ on $D$, and $f, g$ vanish to degrees $3, 4$.  In the
  limit $b^2 \rightarrow 0$, $f, g$ vanish to degrees $3, 5$ and the
  symmetry is enhanced to $\ge_7$.  Note that when $[e_3] \cdot D =
  -1,$ $e_0, e_1$ vanish to degrees $4, 3$ on $D$, giving multiplicities
 $(4, 5)$ along
    $D$ that are enhanced to $(4, 6)$ at points of
  intersection with the remainder of $e_3$, so such models are not
  conventional even before unHiggsing.

As an example of a conventional model of this type, consider on $\F_5$
the $U(1)$ model given by (\ref{eq:two}) with
\begin{eqnarray}
\;[e_3] & = &  2S+ 8F = S + X^{(3)}_{\rm eff}\\
\;[e_2] & = &  4S+ 14F =  2S+ X^{(2)}_{\rm eff}\\
\;[e_1] & = &  6S+ 20F = 2 S+ X^{(1)}_{\rm eff}\\
\;[e_0] & = &  8S+ 26F = 3S+ X^{(0)}_{\rm eff} \,.
\end{eqnarray}
As discussed above, this gives $(3, 4)$ vanishing on the $-5$ curve
$S$ in the $U(1)$ model, enhanced to $(3, 5)$ at points of
intersection with $e_1$.  When $b \rightarrow 0$, the group is
enhanced to $(3, 5)$ on the whole curve $S$, with further enhancement
to $(4, 5)$ at points of intersection with $e_3$.

\subsection{$-6$ curves}
\label{sec:f6}

The situation for $-6$ curves is very similar to that for $-5$
curves.  There is a coupled superconformal theory if $[e_3]$ has positive
intersection with a $-6$ curve, but $[e_3]$ can contain a $-6$ curve
$D$ as a component if $[e_3] \cdot D = -4$, in which case $e_0, e_1,
e_2, e_3$ vanish to orders $3, 2, 2, 1$ on $D$ and the story is
similar to the above.  In this case, however, $e_1$ does not intersect
$D$, so there are no points where this intersection increases the
degree of the singularity.

\subsection{$-7$ curves}
\label{sec:f7}

There are no conventional $U(1)$ configurations of the form
(\ref{eq:two}) where $e_3$ either intersects or contains a curve $D$
of self-intersection $-7$ or below.  The closest to an acceptable
configuration is when $[e_3] \cdot D = -5$, in which case $e_0, e_1,
e_2, e_3$ vanish to orders $3, 3, 2, 1$ on $D$.  This leads to a $(3,
5)$ vanishing of $(f, g)$ on $D$, which is however enhanced to a $(4,
6)$ vanishing at the point where $[e_0]-[D]$ intersects $D$ (of which
there is at least one since $[e_0] \cdot D = -20$).  Any other
combination of intersections leads to a similar singularity.  A
similar problem arises for curves of self-intersection $-8$ or below.

\subsection{The $-3, -2$ non-Higgsable cluster} 
\label{sec:32-cluster}

Finally, we consider the case where $e_3$ contains both a $-3$ curve
$A$ and a  $-2$ curve $B$ that intersects $A$ transversely ($A \cdot B
= 1$).  In this case we find that, at least for some choice of $L,$ a
$(4, 6)$ point is forced at the intersection point between $A$
and $B$.  In particular, we choose $L = -2K$, so that $[e_n] = (n -
4)K$.  From the analysis in \cite{clusters}, we know that a section of
$-4K$ must vanish on $A, B$ to degrees 2, 1 respectively, so
\begin{equation}
[e_0] = 2A + B + X_{\rm eff}^{(0)} \,.
\label{eq:}
\end{equation}
It follows that each of the $e_n$ must contain both $A$ and $B$ as
irreducible components at least once, for $n< 4$,
\begin{eqnarray}
\,[e_1] & = & A + B + X_{\rm eff}^{(1)}\\
\,[e_2] & = & A + B + X_{\rm eff}^{(2)}\\
\,[e_3] & = & A + B + X_{\rm eff}^{(3)}\\
\,[b] & = &  X_{\rm eff}^{(4)} \,.
\end{eqnarray}
Now, consider the degrees of vanishing of the various terms in
(\ref{eq:two}).  While for $b \neq 0$, there are terms in $f$ and $g$
that only vanish to degrees (3, 5) at the intersection of $A$ and $B$
(namely those proportional to $e_0 b^2, e_0e_2 b^2$), when we take $b
\rightarrow 0$, all the remaining non-vanishing terms are of degrees at
least $(4, 6)$ at the intersection point.

This means that in such a situation, while there can exist a 6D
F-theory model with a $U(1)$ gauge symmetry associated with a
nontrivial Mordell-Weil rank, and the Weierstrass coefficients can be
tuned to naively produce a nonabelian $SU(2)$ structure, the resulting
model might have an isolated $(4,6)$ point and hence  be coupled to a
superconformal theory,\footnote{We would like to thank Jim Halverson for discussions on
  this point.  Analogous curves in 4D F-theory models are identified
  and classified in \cite{ghst}.}  
or in other situations \cite{mpt} might have $(4,6)$ singularities
all along one or more curves after 
unHiggsing\footnote{We would like to thank D.\ Park for discussions on
  this point.}, which indicates that these models are at infinite
distance from the interior of the moduli space.
There are many known examples of
base surfaces that contain $-3, -2$ non-Higgsable clusters; a variety
of such examples were constructed in \cite{toric, Martini-WT}.  It
would be interesting to analyze
in detail the structure of $U(1)$ symmetries
that could be tuned over some of these bases.

\section{Implications for 6D and 4D F-theory models} 
\label{sec:remarks}

\subsection{F-theory and supergravity in six dimensions} 
\label{sec:6D}

Six dimensions provides a rich but tractable context in which to study
general aspects of string vacua and quantum supergravity theories.  In
six dimensions, F-theory seems to provide constructions for essentially all known
string vacua, and the space of F-theory vacua matches closely with the
set of potentially consistent quantum supergravity theories
\cite{universality, bound, KMT-I, Seiberg-WT, KMT-II}.  The class of
6D F-theory constructions based on Weierstrass models of elliptically
fibered Calabi-Yau threefolds with section form a single moduli space
of smooth components associated with different bases $B_2$ that are
connected through tensionless string transitions \cite{Seiberg-Witten,
  Morrison-Vafa-II}; recent work has made progress in providing a
global picture of this connected moduli space \cite{KMT-II, clusters,
  Hodge}.  The results of \cite{Braun-Morrison} raised a question of
whether genus-one fibrations without section might constitute a class
of F-theory models that were disconnected from the rest of the
F-theory moduli space.  The picture outlined in this note makes it
clear that in fact the Jacobian fibrations for threefolds without
section fit neatly into the connected moduli space of Weierstrass
models.  Furthermore, this picture sheds light on how $U(1)$ gauge
fields in 6D F-theory models may be understood in the context of the
full moduli space of models.  In \cite{Braun-Morrison,
  Borchmann:2013jwa, Cvetic-Klevers-1, Cvetic-Klevers-2}, a systematic
description was given of the general form for Weierstrass models
containing one, two, and three $U(1)$ fields.  It is known that 6D
models can be constructed with up to eight or more $U(1)$ fields; for
example, as described in \cite{singularities} there are F-theory
constructions on $\P^2$ with an $SU(9)$ tuned on a curve of genus one
that contain an adjoint representation the breaking of which gives
gauge factors $U(1)^8$, and in \cite{Martini-WT} a class of
$\C^*$-bases $B_2$ were found with varying automatic ranks for the
Mordell-Weil group for generic elliptic fibrations; the resulting
threefolds are closely related to the Schoen manifold \cite{mpt}.
One such base in
particular is a generalized
del Pezzo nine over which the generic elliptic fibration has a rank 8
Mordell-Weil group, corresponding to gauge factors $U(1)^8$.  In
\cite{Park-WT}, it was shown from 6D anomaly cancellation arguments
that for a pure abelian theory in 6D with no tensor multiplets
(corresponding to an F-theory model on $\P^2$) the number of $U(1)$
fields is bounded above by $r \leq 17$.  The approach taken in this
paper shows that for a single $U(1)$ factor, it is often possible to
tune the model so that the $U(1)$ can be seen as arising from an
$SU(2)$ or larger nonabelian factor that is Higgsed by VEVs for an adjoint field in a conventional
F-theory model.  It would be interesting to investigate the possible
apparent exceptions to this construction, such as the ones we
encountered with base contains a $-3, -2$ cluster, where the F-theory
model becomes coupled to a superconformal theory.
While in general the construction of higher rank
Mordell-Weil models seems very challenging due to the global nature of
the sections, it would be very interesting to explore when higher rank
abelian models can arise from Higgsed nonabelian gauge symmetries.
This would provide a powerful tool for the construction of general
models with abelian gauge symmetries, since a systematic analysis of
the nonabelian sector is much more straightforward, both in F-theory
and 6D supergravity.  It would also be interesting to explore in more
detail the way in which the basic $SU(2) \to U(1) \to \mathbb{Z}_2$
Higgsing pattern interacts with other nonabelian gauge symmetries
which may be present in a given model.

The existence of an underlying $SU(2)$ for many $U(1)$ gauge factors
also greatly clarifies the set of possible spectra.  The spectrum of
$SU(2)$ theories is quite constrained by anomaly cancellation
\cite{KPT}, which in turn places strong constraints on the spectrum of
possible charges for abelian factors in the 6D supergravity gauge
group.  When an $SU(2)$ factor is tuned on a curve of genus $g$ over a
general base $B_2$ generically the model will include $g$ symmetric
(adjoint) representations and some number of fundamentals.  After
breaking to a $U(1)$, this gives charges 1 and 2, so these are the
only charges expected in generic models.  For specially tuned singular
curves, however, higher representations of $SU(2)$ are possible.  For
example, following the lines of \cite{singularities}, we expect that
an $SU(2)$ on a quintic curve on $P^2$ can carry a 3-symmetric
(4-dimensional) representation when the curve is tuned to have a
triple point of self-intersection.  Group theoretically, this should
correspond to an embedding of $\gsu_2 \oplus \gsu_2\oplus \gsu_2$ in
an $\ge_7$ singularity associated with the triple intersection point.
After breaking the $SU(2)$ to $U(1)$ by an adjoint VEV, this would
give rise to a massless scalar hypermultiplet of charge $\pm$3 under
the $U(1)$.  By the mechanism discussed in this paper, such fields
could then be used to break the $U(1)$ to a discrete $\Z_3$ gauge
symmetry, associated again with a Weierstrass model associated with
the Jacobian of an elliptic fibration with a multisection.  Exploring
the range of possibilities of this type that may be possible for
general representations of $SU(2)$ and higher rank nonabelian groups
on arbitrary curves on general F-theory bases $B_2$ promises to
provide a rich and interesting range of phenomena.  The analysis here
shows that there are strong constraints on the charge spectrum
for $U(1)$ fields in many 6D F-theory models.  These constraints
are stronger than those imposed simply by 6D anomaly
cancellation.  In the spirit of \cite{universality}, it would be
interesting to understand if some of the F-theory constraints on
charge structure could be seen as consistency conditions for the
low-energy 6D supergravity theories with abelian gauge factors.

\subsection{Four dimensions}
\label{sec:4D}

At the level of geometry, the framework developed in this paper should
be valid for Calabi-Yau manifolds of any dimension.  It has not been
shown, however, that all genus-one fibered Calabi-Yau $(n+1)$-folds $X_{n+1}$ that
lack a global section have an associated Jacobian fibration $J_{n+1}$
whose total space is Calabi-Yau
when $n\geq 3$, so it is possible in principle that the analysis
described here can only be applied in a subset of cases where there is
a Jacobian fibration available.  If so, the application to
four-dimensional F-theory constructions would only be relevant in
those cases.  When a Jacobian fibration is available, however, the
analysis of \S\ref{sec:general} should hold: the Jacobian fibrations
of all Calabi-Yau fourfolds with a genus-one fibration but no global
section should fit into the moduli space of Weierstrass models over
complex threefold bases $B_3$, with an explicit description of the
form (\ref{eq:bisection}) when the Jacobian fibration has a bisection.
When the section $e_4 = b^2/4$ is a perfect square, the bisection
becomes a pair of global sections and the Mordell-Weil rank increases
by one.  When $b \rightarrow 0$, the extra section transforms into a
vertical $A_1$ Kodaira type singularity without changing the total
Hodge number $h^{1, 1} (X_4)$.  In many situations, the physics
interpretation of this geometry through F-theory will be the same as
in 6 dimensions: the bisection geometry will be associated with a
discrete $\Z_2$ gauge symmetry that arises from a broken $U(1)$ gauge
field, which in turn can be viewed as coming from an $SU(2)$ gauge
group broken by an adjoint VEV.  Wrapping the 4D theory on a circle
will give distinct vacua, again associated with the Tate-Shafarevich
group and in the M-theory picture with a discrete choice of Calabi-Yau
fourfold with a genus-one fibration but no section.  We also expect a
similar story to hold for higher degree multisections and elliptic
fibrations with higher rank Mordell-Weil group.  In four dimensions,
however, there is additional structure beyond the geometry that can
modify this story.  In particular, G-flux, associated with 4-form flux
of the antisymmetric 3-form potential in the dual M-theory picture,
produces a superpotential that gives masses to many of the scalar
moduli of the Calabi-Yau geometry.  This mechanism can modify the
gauge group and matter spectrum of the theory from that
described purely by the geometry.  At this point a complete
understanding of the role of G-flux in F-theory is still lacking,
despite some recent progress in this direction
\cite{Grimm-hkk,Jockers:2009ti,Marsano:2010ix,Collinucci:2010gz,Braun:2011zm,Marsano:2011hv,Krause:2011xj,Grimm:2011fx,Krause:2012yh,Collinucci:2012as,Intriligator-jmmp}.  We leave the analysis of how the results in this note are
affected by G-flux and the 4D superpotential to further work.
The implications of the generic appearance of an $SU(2)$ (or larger)
nonabelian enhancement for most $U(1)$ vector fields are, however, a question of obvious phenomenological interest.

\vspace*{0.1in}

{\bf Acknowledgements}: We would like to thank Lara Anderson, Volker
Braun, Antonella Grassi, Jim Halverson, and Daniel Park for helpful discussions.
We also thank the organizers and participants of the AMS special
session on ``Singularities and Physics,'' Knoxville, Tennessee, March
2014, during which much of this work was carried out.  This research
was supported by the DOE under contract \#DE-FC02-94ER40818, and by
the National Science Foundation under grant PHY-1307513.

\appendix

\section{Solving the equations determining $I_2$ fibers}
\label{app:solving}

In this appendix, we will explain how to solve the equations 
\eqref{eq:condition1}, \eqref{eq:condition2} which determine the
location of the codimension two $I_2$ fibers by finding the conditions
for the quartic equation to be a square, as in \eqref{eq:square}.  
The first observation is that when the coefficient functions
 $e_0$, $e_1$, \dots $e_4$ are generic, none of them will vanish
at any of the solutions to \eqref{eq:square}.

In the case of $e_4$, if $e_4$ vanishes at a solution then $u=0$
is one of the double roots so $e_3$ must also vanish.  For the remaining
root to be double, we also need $e_1^2=4e_0e_2$ to vanish, but now we
have three conditions on the base and the solutions are in codimension two.
The case of $e_0$ is similar: if it vanishes, then $e_1$ and $e_3^2-e_2e_4$
would both also have to vanish.

In the case of $e_3$ vanishing, $\beta$ would need to vanish and then
the equation  would take the form $e_4((e_2/2e_4)u^2+v^2)^2$.  Again
we get three condtions: $e_3=0$, $e_1=0$, and $e_2^2=4e_0e_4$ which is
of too large a codimension to be generic.  The case of $e_1$ is similar.

Finally, in the case of $e_2$ vanishing, we have an equation of the
form $e_4(-(e_3^2/8e_4^2)u^2+(e_3/2e_4)uv + v^2)^2$, and this implies the
additional conditions $e_3^3=-8e_4^2e_1$ and $e_3^4=64e_4^3e_0$.  Once again
we have three conditions and this is not possible.

Now we turn to the solution of \eqref{eq:condition1}, \eqref{eq:condition2}.
As in \S\ref{sec:I2}, 
the first step is to introduce an auxiliary variable $p$, and
to express the solutions as the common zeros of two auxiliary polynomials
\begin{align}
\Phi_1 &:= p^4-8e_2e_4p^2+16e_2^2e_4^2-64e_0e_4^3\\
\Phi_2 &:= p^3 - 4e_2e_4p + 8 e_1e_4^2 &
\end{align}
(together with the equation $e_3=p$).  From this, we can form additional
polynomials which must vanish on the solution, roughly following the
Gr\"obner basis algorithm (but allowing division by $e_1$, $e_2$ or $e_4$, 
which are known
not to vanish on solutions).  This gives the following sequence of
polynomials:
\begin{align}
\Phi_3 &:= (-\Phi_1+p\Phi_2)/4e_4 = 
e_2p^2+2e_1e_4p-4e_2^2e_4+16e_0e_4^2\\
\Phi_4 &:= (-e_2\Phi_2+p\Phi_3)/2e_4 
= e_1p^2 + 8e_0e_4p-4e_1e_2e_4 \\
\Phi_5 &:= (e_1\Phi_3-e_2\Phi_4)/2e_4 =
(e_1^2-4e_0e_2)p + 8 e_0e_1e_4 \\
\Phi_6 &:= ((4e_0e_2-e_1^2)\Phi_4+pe_1\Phi_5)/4e_2e_4
= 8e_0^2p-e_1(4e_0e_2-e_1^2) \\
\Phi_7 &:= (8e_0^2\Phi_5 + (4e_0e_2-e_1^2)\Phi_6)/e_1
= 64e_0^3e_4 - (4e_0e_2-e_1^2)^2 \,.
\end{align}
The variable $p$ has been eliminated from $\Phi_7$, so the equation
$\Phi_7=0$ gives the condition for a $p$ to exist (this is
\eqref{eq:condition}).  The equation $\Phi_5=0$ can then be solved
for $p$; this gives \eqref{eq:commonroot}.

\end{document}